\RequirePackage[l2tabu, orthodox]{nag}
\documentclass[journal=nalefd, manuscript=article]{achemso}

\usepackage[T1]{fontenc}
\usepackage{gensymb} %\micro, \ohm, \degree, \celsius
\usepackage[seperr,load={}]{siunitx}
\usepackage{lmodern}
\usepackage{amsmath}
\usepackage{amssymb}
\usepackage{graphicx}
\usepackage{color}
\usepackage{soul}
\usepackage{nccmath}
\usepackage{lineno} 
\usepackage[normalem]{ulem}
\usepackage{hyperref}
\usepackage{url}
\usepackage{gensymb}
\usepackage[utf8]{inputenc}
\usepackage{comment}
\usepackage{array,multirow}
\usepackage{afterpage}
\usepackage{subfiles}
\usepackage{indentfirst}

\usepackage{xargs}
\usepackage[pdftex,dvipsnames]{xcolor}
\usepackage[colorinlistoftodos,prependcaption,textsize=footnotesize]{todonotes}
\newcommandx{\unsure}[2][1=]{\todo[linecolor=red,backgroundcolor=red!25,bordercolor=red,#1]{#2}}
\newcommandx{\change}[2][1=]{\todo[inline,linecolor=blue,backgroundcolor=blue!25,bordercolor=blue,#1]{#2}}
\newcommandx{\info}[2][1=]{\todo[inline,linecolor=OliveGreen,backgroundcolor=OliveGreen!25,bordercolor=OliveGreen,#1]{#2}}
\newcommandx{\improvement}[2][1=]{\todo[inline,linecolor=Plum,backgroundcolor=Plum!25,bordercolor=Plum,#1]{#2}}
\newcommandx{\thiswillnotshow}[2][1=]{\todo[disable,#1]{#2}}
\newcommandx{\todosection}[2][1=]{\todo[inline,size=\large,#1]{#2}}
\usepackage{color}

\title{Ballistic one-dimensional holes with strong g-factor anisotropy in germanium} 
\author{R. Mizokuchi}
\author{R. Maurand}
\author{F. Vigneau}
\affiliation{Universit\'e Grenoble Alpes \& CEA, INAC-PHELIQS, F-38000 Grenoble, France}
\author{M. Myronov}
\affiliation{Department of Physics, University of Warwick, Coventry CV4 7AL, United Kingdom}
\author{S. De Franceschi}
\affiliation{Universit\'e Grenoble Alpes \& CEA, INAC-PHELIQS, F-38000 Grenoble, France}
\email{silvano.defranceschi@cea.fr}
\begin{document}

\begin{abstract}
We report experimental evidence of ballistic hole transport in one-dimensional quantum wires  gate-defined in a strained SiGe/Ge/SiGe quantum well. At zero magnetic field, we observe conductance plateaus at integer multiples of $2e^2/h$ . At finite magnetic field, the splitting of these plateaus by Zeeman effect reveals largely anisotropic g-factors, with absolute values below 1 in the quantum-well plane, and exceeding 10 out of plane.  This g-factor anisotropy is consistent with a heavy-hole character of the propagating valence-band states, in line with a predominant confinement in the growth direction. Remarkably, we observe quantized ballistic conductance in device channels up to 600 nm long. These findings mark an important step towards the realization of novel devices for applications in quantum spintronics.

\end{abstract}

\maketitle
Quantum spintronics is an active research field aiming at the development of semiconductor quantum devices with spin-based functionality \cite{Awschalom2007Challenges}. This field is witnessing an increasing interest in exploiting the spin degree of freedom of hole spin states, which can present a strong spin-orbit (SO) coupling, enabling electric-field driven spin manipulation \cite{Nadj-Perge2012Spectroscopy,Pribiag2013Electrical}, and a reduced hyperfine interaction, favoring spin coherence \cite{Gerardot2008Optical,Bulaev2007Electric,Fischer2008Spin}.  

Efficient electric-dipole spin resonance was recently demonstrated for hole spins confined in silicon quantum dots \cite{Maurand2016CMOS,Ono2017Hole}. Even faster manipulation should be possible in germanium, where holes have stronger SO coupling \cite{Watzinger2018Hole}. Germanium is also known to form low-resistive contacts to metals, owing to a Fermi-level pinning close to the germanium valence band. This property can lead to interesting additional opportunities, such as the realization of hybrid superconductor-semiconductor devices \cite{DeFranceschi2010Hybrid} (e.g. Josephson field-effect transistors \cite{Clark1980Feasibility}, gatemons\cite{deLange2015Realization,Casparis2016Gatemon}, and topological superconducting qubits based on Majorana fermions \cite{Lutchyn2017Realizing,Maier2014Majorana}, for which the concomitant presence of strong SO coupling would play a key role).

Experimental realizations of Ge-based nanoelectronic devices have so-far relied primarily on bottom-up nanostructures: Ge/Si core/shell nanowires (NWs) \cite{Lu2005One,Kotekar2017Quasiballistic, Xiang2006Ge, Su2016High}, SiGe self-assembled quantum dots \cite{Katsaros2011Observation, Ares2013SiGe}, and Ge hut NWs \cite{Zhang2012Monolithic}. Following recent progress in SiGe epitaxy, SiGe/Ge/SiGe quantum-well heterostructures embedding a high-mobility two-dimensional hole gas have become available \cite{Maksym2014extremely,Shi2015Spinless,laroche2016magneto}, providing a new attractive option for the realization of quantum nanoelectronic devices  \cite{Gul2017Quantum,Gul2018Self,Hendrickx2018Gate}. 

Here we report the fabrication and low-temperature study of devices comprising a gate-tunable, one-dimensional (1D) hole channel with a gate-defined length varying between 100 and 900 nm. We reveal the ballistic 1D nature of hole transport through measurements of conductance quantization. By measuring the Zeeman splitting of the conductance plateaus in a magnetic field, $\vec{B}$,  applied along different directions, we find a strong g-factor anisotropy consistent with a dominant heavy-hole (HH) character of the 1D subbands. 

%In order to avoid possible problems of parallel conduction through modulation-doping layers \cite{MOrris} and to reduce scattering from remote dopants 

The devices were fabricated from a nominally undoped heterostructure consisting of a pseudomorphically strained, 22-nm thick Ge quantum well (QW) confined by $\mathrm{Si_{0.2}Ge_{0.8}}$ barriers, i.e. a relaxed $\mathrm{Si_{0.2}Ge_{0.8}}$ buffer layer below, and a 72-nm-thick $\mathrm{Si_{0.2}Ge_{0.8}}$ layer above, capped by 2 nm of low-temperature-grown Si. The heterostructure was grown by reduced pressure chemical vapor deposition on a Si(001) wafer (See Ref. \cite{Myronov2014Extremely} and details therein). 

At low temperature,  the Ge QW is carrier free, and hence insulating, due to the intentional absence of doping. A two-dimensional hole gas with a mobility of $1.7 \times \SI{e5}{cm^2/V.s}$ and a hole density of $\sim \SI{e11}{cm^{-2}}$ can be electrostatically induced by means of a negatively biased top gate electrode (for more details see Supplementary Information). 

The device layout consists of a mesa structure defined by optical lithography and reactive ion etching with $\mathrm{Cl_{2}}$ gas. Two platinum contact pads, to be used as source and drain electrodes, are fabricated on opposite sides of the mesa. Platinum deposition is carried out after dry-etch removal of the SiGe overlayer followed by a two-step surface cleaning process to eliminate the native oxide (wet HF etching followed by Ar plasma bombardment in the e-beam evaporator). We obtain contact resistances of the order of few $\si{k \Omega}$. An $\mathrm{ Al_{2} O_{3}}$ 30-nm thick gate oxide layer is deposited by atomic layer deposition at $250\,\mathrm{C}\degree$. Ti/Au top-gate electrodes are finally defined using e-beam lithography and e-beam metal deposition: a central gate extending over the mesa is designed to induce the accumulation of a conducting hole channel between the source to the drain contact; two side gates, to be operated in depletion mode, create a tunable 1D constriction in the channel oriented along the [100] direction. We have varied the geometry of the side gates in order to explore gate-defined 1D hole wires with different lengths. Here we present experimental data for two devices, one with a short ($\sim100$ nm) and one with a long ($\sim600$ nm) constriction (see Figs. ~\ref{Fig:QPC} (a) and (b), respectively). 

All magnetotransport measurements were done at $270\,$mK in a $\mathrm{^{3}He}$ cryostat equipped with a superconducting magnet. Figure~\ref{Fig:QPC} (c) shows a data set for a device, labelled D1, nominally identical to the one shown in Fig.~\ref{Fig:QPC} (a).  The differential conductance, $G$, measured at dc source-drain bias voltage $V_{ds} = 0$, is plotted as a function of $V_{sg}$ for magnetic fields, $\vec{B}$, perpendicular to the QW plane and varying from 0 to 0.5 T. In our experiment, $G$ was directly measured using standard lock-in detection with a bias-voltage modulation $\delta V_{sd} = 10\,\si{\mu V}$ at $36.666\,$Hz. In addition, $G$ was numerically corrected to remove the contribution from all series resistances ($\sim20$ k$\Omega$), i.e. the resistances of the measurement circuit, the source and drain contacts, and the two-dimensional hole gas.  

$G$ exhibits clear quantized  plateaus in steps of $2e^{2}/h$, where $e$ is the electron charge and $h$ is the Plank constant. This finding is consistent with the results of a recently published independent work carried out on a similar SiGe heterostructure \cite{Gul2017Quantum}. Applying an out-of-plane magnetic field lifts the spin degeneracy of the 1D subbands, resulting in plateaus at multiples of $e^{2}/h$. These plateaus underpin the formation of spin-polarized subbands. They emerge at relatively small magnetic fields, of the order of a few hundred mT, denoting a large out-of-plane $g$-factor as expected in the case of a predominant HH character. 

We measured several devices with side-gate lengths, $L_g$, ranging from 100 nm (as in Fig.\ \ref{Fig:QPC} (a)) to 900 nm. The $G(V_{sg})$ measurements shown in Fig.~\ref{Fig:QPC}(d) were taken on a device with $L_g \approx 600$ nm, labelled as D2 and nominally identical to the one shown in Fig. \ref{Fig:QPC}(b). Remarkably, these measurements demonstrate that clear conductance quantization can be observed also in relatively long channels largely exceeding 100 nm.  

We note that a shoulder at $G \sim 0.7 \times 2e^2/h$ is visible in the $B=0$ traces of both Fig. ~\ref{Fig:QPC} (c) and (d). This feature, which is highlighted in the respective insets, corresponds to the so-called 0.7 anomaly. Discovered and widely studied in quantum point contacts defined in high-mobility two-dimensional electron systems  \cite{vanWees1988Quantized,Thomas1996Possible,Kristensen2000Bias,Cronenwett2002Low,Danneau2008Structure,Komijani2013Origins}, and more recently observed also in semiconductor nanowires  \cite{Heedt2016Ballistic,Saldana2018Supercurrent},  the interpretation of this phenomenon remains somewhat debated \cite{Micolich2011Tracking,Iqbal2013Odd,Bauer2013Microscopic,Brun2014Wigner,Iagallo2015Scanning}.

\begin{figure}[p!]
\centering
\includegraphics[width=1\linewidth]{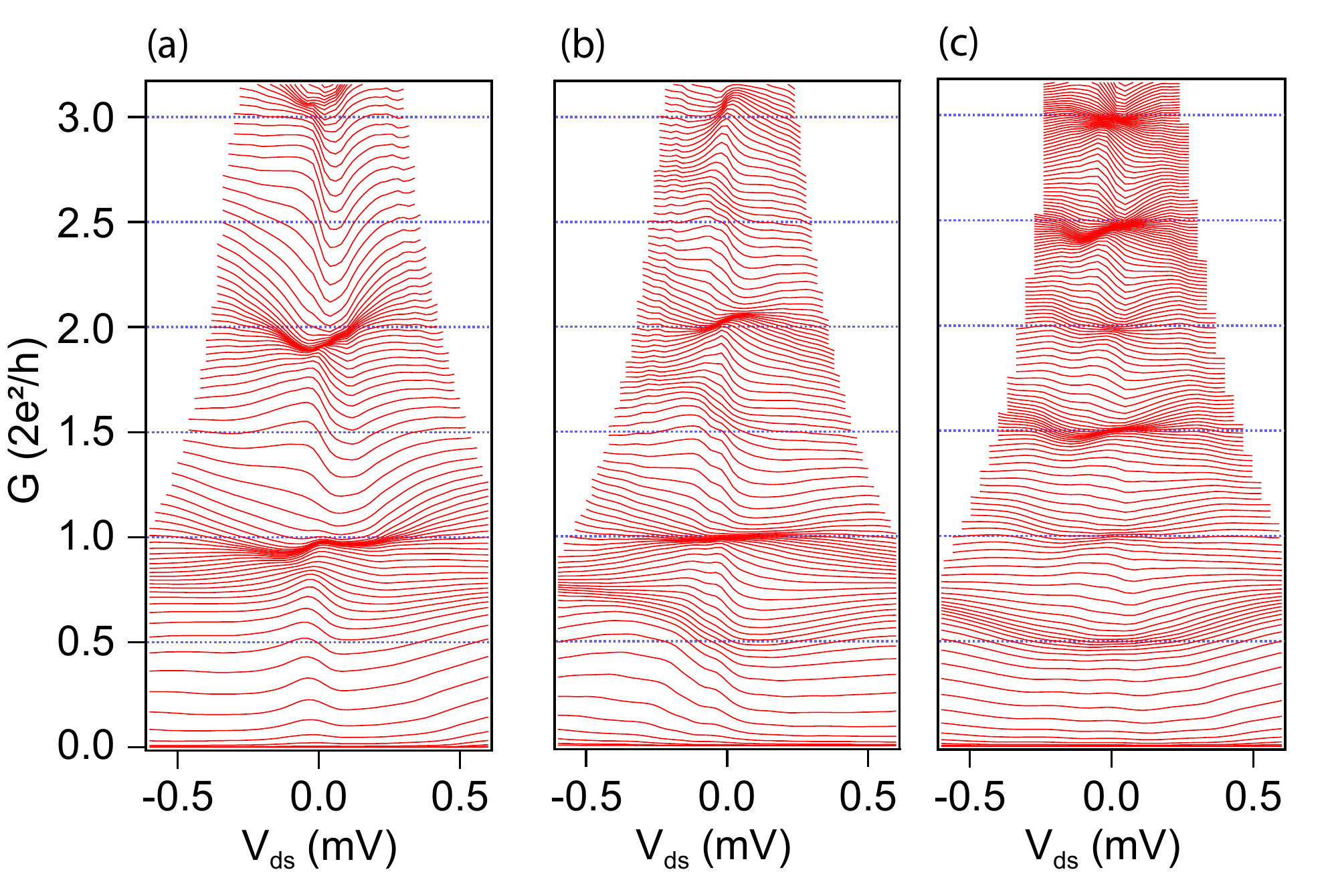}
	\caption{\label{Fig:WaterFall} Waterfall plots of differential conductance, $G$, as a function of source-drain bias, $V_{ds}$, at different values of side-gate voltage $V_{sg}$ (gate step: $5$\,mV). The three plots were taken on device D1 at different out-of-plane magnetic fields: (a) 0\,T, (b) 0.3\,T and (c) 0.5\,T. 
The spanned  $V_{ds}$ range varies with $V_{sg}$, and hence with $G$. This follows from the procedure used to take into account the effect of the series resistance, $R_S$. In this procedure, we assumed $R_S$ to be monotonically increasing with the current $I_{sd}$ flowing across the device.  This assumption was motivated by the need to account for non-linearities in the series resistance coming primarily from the source/drain contacts to the two-dimensional hole gas. At $V_{sd} = 0$, $R_S$ is a constant all over the spanned $V_{sg}$ range. At finite $V_{sd}$, $R_S$ varies with $V_{sg}$ due to the $V_{sg}$ dependence of $G$. As a result, the corrected $V_{ds}$ range tends to decrease when lowering $V_{sg}$, and hence increasing $G$.} %\GRN{$V_\mathrm{tg} = -2.5$ V}}
\end{figure}

To further confirm the 1D nature of the observed conductance quantization, we present in Figs. \ref{Fig:WaterFall} (a)-(c) waterfall plots of the non-linear $G$($V_{ds}$) at three different perpendicular magnetic fields ($B$ = 0, 0.3, and 0.5 T, respectively) for device D1. Clear bunching of the $G$($V_{ds}$) is observed around $V_{ds}=0$ for gate voltages corresponding to the quantized conductance plateaus of Fig.~\ref{Fig:QPC}(c). With magnetic field applied, the first plateau at $G = e^2/h$ begins to appear at $B = 0.3 \ $T and is fully formed at $B = 0.5 \ $T. At $B=0$, a zero-bias $dI/dV$ peak can seen in correspondence of the $0.7$ structure, in line with previous observations \cite{Cronenwett2002Low}.   
    
The well-resolved spin splitting of the 1D subbands enables a quantitative study of the hole g-factors. To investigate the g-factor anisotropy, we applied $\vec{B}$ not only along the z axis, perpendicular to the substrate plane, but also along the in-plane directions x and y, indicated in Figure~\ref{Fig:QPC} (a). To change the $\vec{B}$ direction, the sample had to be warmed up, rotated, and cooled down multiple times. Thermal cycling did not modify significantly the device behavior, except for the value of threshold voltage on the channel gate for the activation of hole conduction in the Ge QW (this voltage is sensitive to variations in the static charges on the sample surface). 

\begin{figure}[p!] 
	\centering
	\includegraphics[width=1\linewidth]{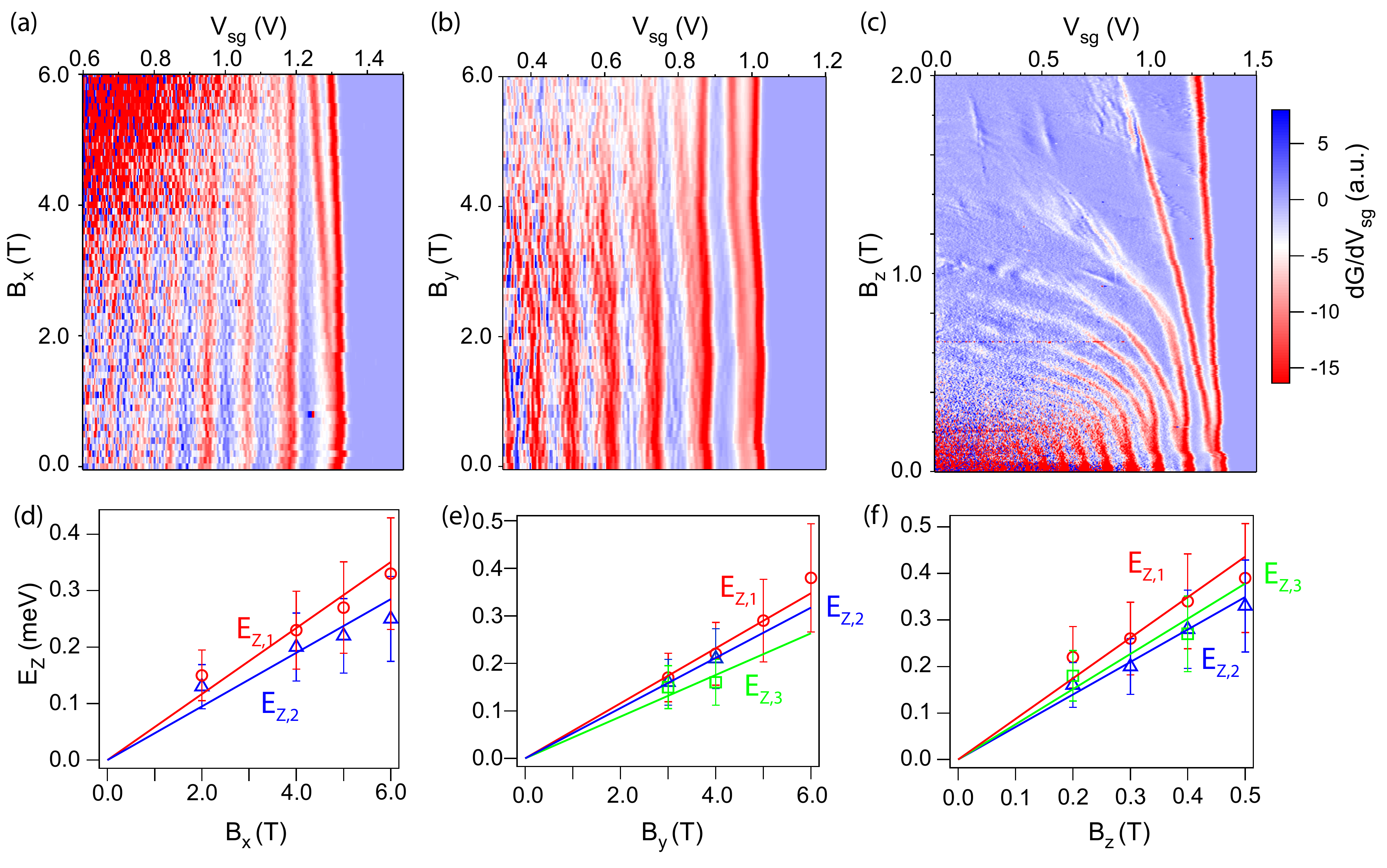}
	\caption{\label{Fig:Map}(a)-(c) Numerical derivative of $G$ with respect to $V_{sg}$ as a function of $V_{sg}$ and magnetic field applied along the $x$ (a), $y$ (b) and $z$ (c) directions (data from device D1). (d)-(f) Zeeman splittings $E_{\mathrm{Z},n} = \left| E_{n,\uparrow} - E_{n,\downarrow}\right|$ as a function of magnetic field along the $x$ (d), $y$ (e) and $z$ (f) directions. Red, blue, and green open symbols  correspond to the first, second, and third spin-split subbands, respectively. The g factors for each subband are obtained from the slope of the linear fits to the Zeeman relation $E_{\mathrm{Z},n}(B)$ (solid lines). The results are given in Table \ref{Table:gfactor}. %\GRN{Data for each magnetic field direction is taken at $V_{\mathrm{tg}} = \text{(x) }-2.4 \text{ V, } \text{(y) and (z) }-2.5 \text{ V}$, respectively.}}
}
\end{figure}

Figures ~\ref{Fig:Map} (a), (b) and (c) show the $B$-evolution of the trans-conductance $dG/dV_{sg}$ as a function of $V_{sg}$, with $\vec{B}$ applied along x, y and z, respectively. The data refer to device D1. In these color maps, the blue regions, where $dG/dV_{sg}$ is largely suppressed, correspond to the plateaus of quantized conductance. On the other hand, the red ridges of enhanced $dG/dV_{sg}$  correspond to the conductance steps between consecutive plateaus, which occur every time the edge of a 1D subband crosses the Fermi energy of the leads. At finite $B$, the red ridges split, following the emergence of new conductance plateaus at odd-integer multiples of $e^2/h$. 
Upon increasing $B$, the splitting in $V_{sg}$ increases proportionally to the Zeeman energy $E_{\mathrm{Z},n} = \left| E_{n,\uparrow} - E_{n,\downarrow}\right|$, where $E_{n,\sigma}$ is the energy of the 1D subband with spin polarization $\sigma$ and orbital index $n$. 

For an in-plane $B$, either along $x$ or $y$, the splitting becomes clearly visible only above approximately 2 T. As a result, the explored $B$ range extends up to 6 T. For a perpendicular field, the Zeeman splitting is clearly more pronounced being visible already around 0.2 T. This apparent discrepancy reveals a pronounced g-factor anisotropy, with a g-factor along the z-axis, $g_z$, much larger than the in-plane g-factors, $g_x$ and $g_y$. Such a strong anisotropy is expected in the case of two-dimensional hole states with dominant HH character, corroborating the hypothesis of a dominant confinement in the $z$ direction, which is imposed by the QW heterostructure. 

Besides causing the Zeeman splitting of the 1D subbands, the applied $\vec{B}$ has an effect on the orbital degree freedom of the hole states. The effect is relatively weak in the case of an in-plane $B$ because the magnetic length, inversely proportional to $\sqrt{B}$, gets as small as the QW thickness only for the highest $B$ values spanned in Figs. ~\ref{Fig:Map}(a) and ~\ref{Fig:Map}(b). On the contrary, the relatively weak lateral confinement imposed by the side gates leaves room for a pronounced $B$-induced orbital shift. This manifests in Fig. ~\ref{Fig:Map} (c) as an apparent bending of the $dG/dV_{sg}$ ridges towards more negative gate voltages.

In order to quantitatively estimate the observed Zeeman splittings, and the corresponding g-factors, we performed bias-spectroscopy measurements of $dG/dV_{sg}$ as a function of $V_{ds}$ and $V_{sg}$ at different magnetic fields. In these measurements, the $dG/dV_{sg}$ ridges form diamond-shape structures from which we extract the Zeeman energies, as well as the lever-arm parameters relating $V_{sg}$ variations to energy variations. 
Representative $dG/dV_{sg}$($V_{ds}$,$V_{sg}$) measurements, and a description of the well-known procedure for the data analysis are given in Supplementary Information. Interestingly, for a given $B$, both the Zeeman energy and the lever-arm parameter can vary appreciably from subband to subband.  

Figures~\ref{Fig:Map} (d), (e), and (f) present the estimated $E_{\mathrm{Z},n}$ values as a function of $B$, for the first few subbands, and for the three $B$ directions. Linear fitting to $E_{\mathrm{Z},n} =g_n \mu_{\mathrm{B}}B$  yields the Land\'e $g$-factors, $g_{x,n}$, $g_{y,n}$, and $g_{z,n}$ for the three perpendicular directions.  The extracted $g$-factors for the device D1 are listed in Table~\ref{Table:gfactor}. We have included also  the $g_{z,n}$ values obtained from another device (D3) with $L_g = 100$ nm. 

For device D1 (D3), the perpendicular g-factor ranges between 12.0 (10.4) and 15.0 (12.7), while the in-plane one is much smaller, varying between 0.76 and 1.00, with no significant difference between $x$ and $y$ directions. 
A large in-plane/out-of-plane anisotropy in the g-factors is consistent with the hypothesis of a dominant HH character. In fact, in the limit of vanishing thickness, the lowest subbands of a Ge QW should approach pure HHs with $g_x \approx g_y \approx 0$ and  
$g_z=6\kappa+27q/2=21.27$, where $\kappa$ and $q$ are the Luttinger parameters ($\kappa = 3.41$ and $q=0.06$ for Ge).

%For the first subband we find $g_{z,1}=15.0 \pm 2.3$ and $g_{z,1}=12.7 \pm 2.2$ for device D1 and D3, respectively. 

%These values are very close to the perpendicular 
%g-factor of a pure HH state in two-dimensional Ge, i.e. in the limit of zero quantum-well thickness, for which we expect $g_z=6\kappa+27q/2=21.27$, where $\kappa$ and $q$ are the Luttinger parameters ($\kappa = 3.41$ and $q=0.06$ for Ge). 
%The measured in-plane g-factors are much smaller, in line with the hypothesis of a dominant HH character, which should result in vanishing in-plane g-factors. The in-plane anisotropy is weak, with a discrepancy between $x$ and $y$ components close to the experimental uncertainty ($g_{x,1}=1.40 \pm 0.08$ and $g_{y,1}=1.39 \pm 0.15$). 

In the investigated SiGe QW heterostructure, the HH nature of the first 2D subbands is enhanced by the presence of a biaxial compressive strain in the Ge QW,  increasing by $\sim40\ \si{meV}$ the energy splitting with the first light-hole (LH) subbands \cite{Failla2015Narrow}. The creation of a 1D constriction does not introduce a significant HH-LH mixing because confinement remains dominated by the QW along the growth axis ($z$). From a measured energy spacing of around 0.65 meV between the first and the second 1D subband (see Supplementary Information), we estimate that the hole wavefunctions of the first subband have a lateral width (along $y$) of approximately 80 nm, which is an order of magnitude larger than the wavefunction extension along $z$.   

The results summarized in Table~\ref{Table:gfactor} suggest a slight tendency of the g-factors to decrease with  the subband index. 
This trend is consistent with the results of earlier experiments with both electron \cite{martin2008enhanced,Heedt2016Ballistic} and hole \cite{Daneshvar1997Enhanced,Danneau2006Zeeman,Srinivasan2016Electrical} quantum point contacts. A possible explanation is that the exchange interaction increases the g-factor in the low-density limit \cite{Thomas1996Possible,Wang1996Spin}. Yet hole g-factors in quantum point contacts depend also on a complex interplay of spin-orbit coupling, applied magnetic field, and electrostatic potential landscape \cite{miserev2017dimensional,Miserev2017Mechanisms}. Acquiring a deep understanding of the g-factors reported here would require more extensive and sophisticated experiments together with a nontrivial theoretical analysis, which goes well beyond the scope of the present work.   

In conclusion, we have demonstrated ballistic hole transports in 1D quantum wires gate-defined in a $\mathrm{Ge/Si_{0.2}Ge_{0.8}}$ heterostructure. The ballistic regime is observed for wires up to $600$\,nm long. By investigating the Zeeman splitting of the quantized conductance steps we find that out-of-plane g-factors are an order of  magnitude larger than the in-plane ones, denoting a pronounced HH character. This can be ascribed to the dominant confinement along the growth axis and to the compressive biaxial strain in the Ge QW. 
The observation of ballistic 1D hole transport in remarkably long channels and large out-of-plane g-factors holds special promise for the development of devices with spin-related functionality. In principle, the fabrication of these devices could be implemented in an industry-standard fab line with the possibility of monolithic integration with conventional silicon electronics. 

\section{Supplementary Information}
Additional experimental data from a gated Hall-bar device, providing information of the transport properties of the two-dimensional hole gas. Additional data from another 1D-wire device (D3). Description of the procedure to extract energy spacings in a 1D channel. 

\section{Acknowledgement}
We acknowledge financial support from the Agence Nationale de la Recherche, through the TOPONANO project. 

%\begin{figure}[t]
%\begin{center}
%\includegraphics[width=1\linewidth]{FigHallBar.pdf}
%\end{center}
%\caption{\label{Fig:Hallbar}   }
%\end{figure}
%\bibliographystyle{unsrt}
\providecommand{\latin}[1]{#1}
\makeatletter
\providecommand{\doi}
  {\begingroup\let\do\@makeother\dospecials
  \catcode`\{=1 \catcode`\}=2 \doi@aux}
\providecommand{\doi@aux}[1]{\endgroup\texttt{#1}}
\makeatother
\providecommand*\mcitethebibliography{\thebibliography}
\csname @ifundefined\endcsname{endmcitethebibliography}
  {\let\endmcitethebibliography\endthebibliography}{}

\afterpage{\clearpage}

\begin{figure}[p!] 
	\centering
	\includegraphics[width=1\linewidth]{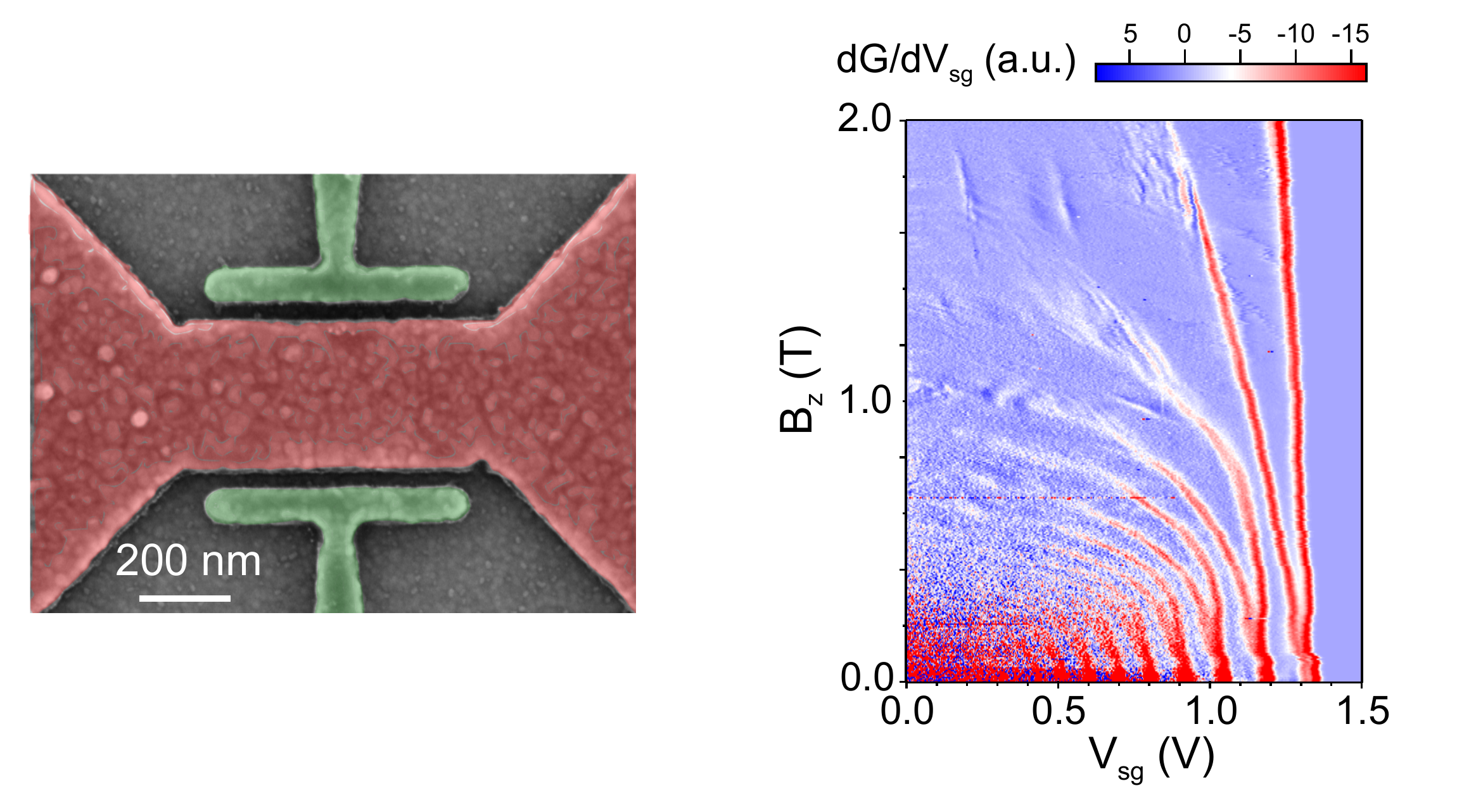}
	\caption*{figure for abstract}
\end{figure}

\begin{figure}[p!]
\centering
\includegraphics[width=1\linewidth]{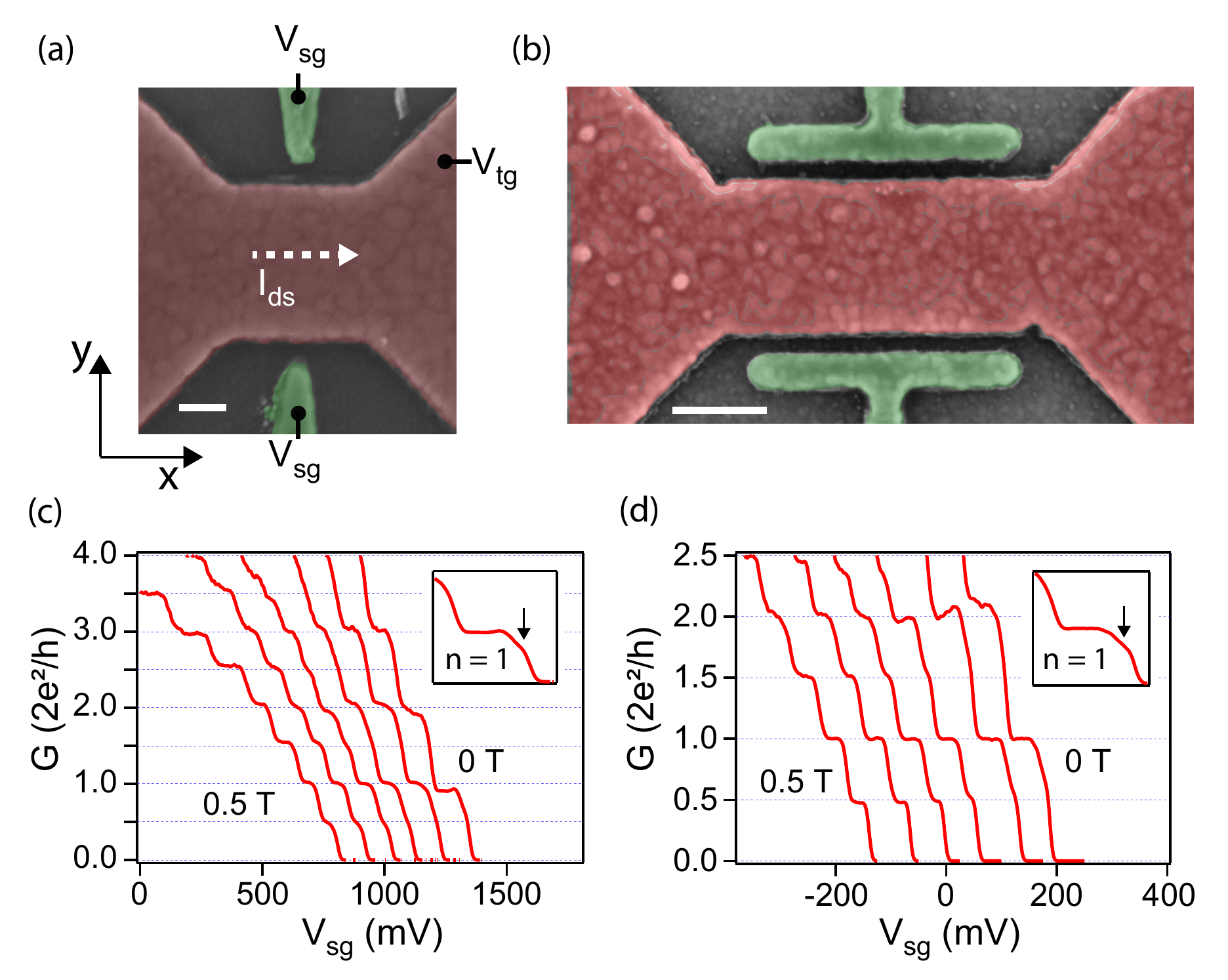}
\caption{\label{Fig:QPC} (a) and (b) False color scanning electron micro-graphs of representative  devices. Scale bars: 100\,nm (a) and  200\,nm (b). Gate voltages $V_{tg} < 0$ and $V_{sg}>0$ are applied to the channel gate (colorized in red) and the two side gates (colorized in green), respectively. Current $I_{ds}$ flows in Ge QW under the channel gate along the $x$ direction. To enable that, the channel gate extends of all the way to the source/drain contact pads, which are located about 15 $\mu$m away from nanowire constriction, i.e. outside of the view field in (a) and (b).  (c) and (d) Measurements of zero-bias conductance $G$ as a function of $V_{sg}$ at different perpendicular magnetic fields, $B_z$, from 0 to 0.5 T (step: 0.1 T). Data in (c) ((d)) refer to device D1 (D2), which is nominally identical to the one shown in (a) ((b)).  
%f$V_{\mathrm{tg}} =$ -2.5 V and in (d) D2 for $V_{\mathrm{tg}} =$  -750 mV}. 
In both cases we observe clear conductance quantization and the lifting of spin degeneracy at finite field. 
Conductance has been rescaled to remove the contribution of a series resistance $R_S$ slightly varying with $B_z$ between 22 and 24 k$\Omega$. The different traces are laterally offset for clarity. Insets: Zoom-in of the 0.7 anomaly (indicated by an arrow) at zero magnetic field.}
\end{figure}

\clearpage
\begin{table}[p!]
      \centering
	\begin{tabular}{|c|c|c|c|c|}\hline
		\multirow{1}{*}{ } & \  & {$g_{1}$} & {$g_{2}$} & {$g_{3}$} \\\hline
		\multirow{3}{*}{D1}  
		& $B_x$ & {1.00 $\pm$ 0.15 } & {0.82 $\pm$ 0.12 } & -  \\\cline{2-5}
		& $B_y$ & {1.00 $\pm$ 0.15 } & { 0.91 $\pm$ 0.19} & { 0.76 $\pm$ 0.16}  \\\cline{2-5}
		& $B_z$ & {15.0 $\pm$ 2.3} & {12.0 $\pm$ 1.8} & {13.0 $\pm$ 2.8} \\\hline
		\multirow{1}{*}{D3} & $B_z$ & {12.7 $\pm$ 2.2} & { 11.8 $\pm$ 1.8} & {10.4 $\pm$ 1.6}\\\hline
	\end{tabular}
\caption{\label{Table:gfactor} This table summarizes the results of g-factor measurements on device D1 and D3. These g-factors are obtained from the slope of the linear fits in Fig. \ref{Fig:Map} (d)-(f)) and Fig. \ref{Fig:Dev3} (c).}
\end{table}
\clearpage

%\afterpage{
%\begin{table}[ht!]
%      \centering
%	\begin{tabular}{|c|c|c|c|c|c|}\hline
%		\multirow{1}{*}{ } & \  & {$g^{*}_{1}$} & {$g^{*}_{2}$} & %{$g^{*}_{3}$} & {$g^{*}_{4}$}\\\hline
%		\multirow{3}{*}{D1}  
%		& $B_x$ & {1.00 $\pm$ 0.15 } & {0.82 $\pm$ 0.12 } & - & - %\\\cline{2-6}
%		& $B_y$ & {1.00 $\pm$ 0.15 } & { 0.91 $\pm$ 0.19} & { 0.76 $\pm$ 0.16} & - \\\cline{2-6}
%		& $B_z$ & {15.0 $\pm$ 2.3} & {12.0 $\pm$ 1.8} & {13.0 $\pm$ 2.8} %& -\\\hline
%		\multirow{1}{*}{D3} & $B_z$ & {12.7 $\pm$ 2.2} & { 11.8 $\pm$ %1.8} & {10.4 $\pm$ 1.6} & {6.2 $\pm$ 0.8}\\\hline
%	\end{tabular}
%\caption{\label{Table:gfactor} This table summarizes the results of g-factor measurements on device D1 and D3. These g-factors are obtained from the slope of the linear fits of the Zeeman splitting as a function of magnetic field amplitude (see Fig. \ref{Fig:Map} (d)-(f)) and Fig. \ref{Fig:Dev3} in Supplementary Information).}
%\end{table}
%}

\afterpage{\clearpage
\pagebreak}
\appendix

\section{Supplementary materials: Ballistic one-dimensional hole transport in Ge/SiGe heterostructure}
%%%%%%%%%%%%%%%%%%%%%%%% Hall measurements %%%%%%%%%%%%%%%%%%%%%%%%

\setcounter{figure}{0}
\setcounter{table}{0}

\renewcommand{\theequation}{S\arabic{equation}}
\renewcommand{\thefigure}{S\arabic{figure}}
\renewcommand{\thetable}{S\arabic{table}}
\renewcommand{\bibnumfmt}[1]{[S#1]}
\renewcommand{\citenumfont}[1]{S#1}

\section{Hall measurements}
\begin{figure}[p!]
	\centering
	\includegraphics[width=1\linewidth]{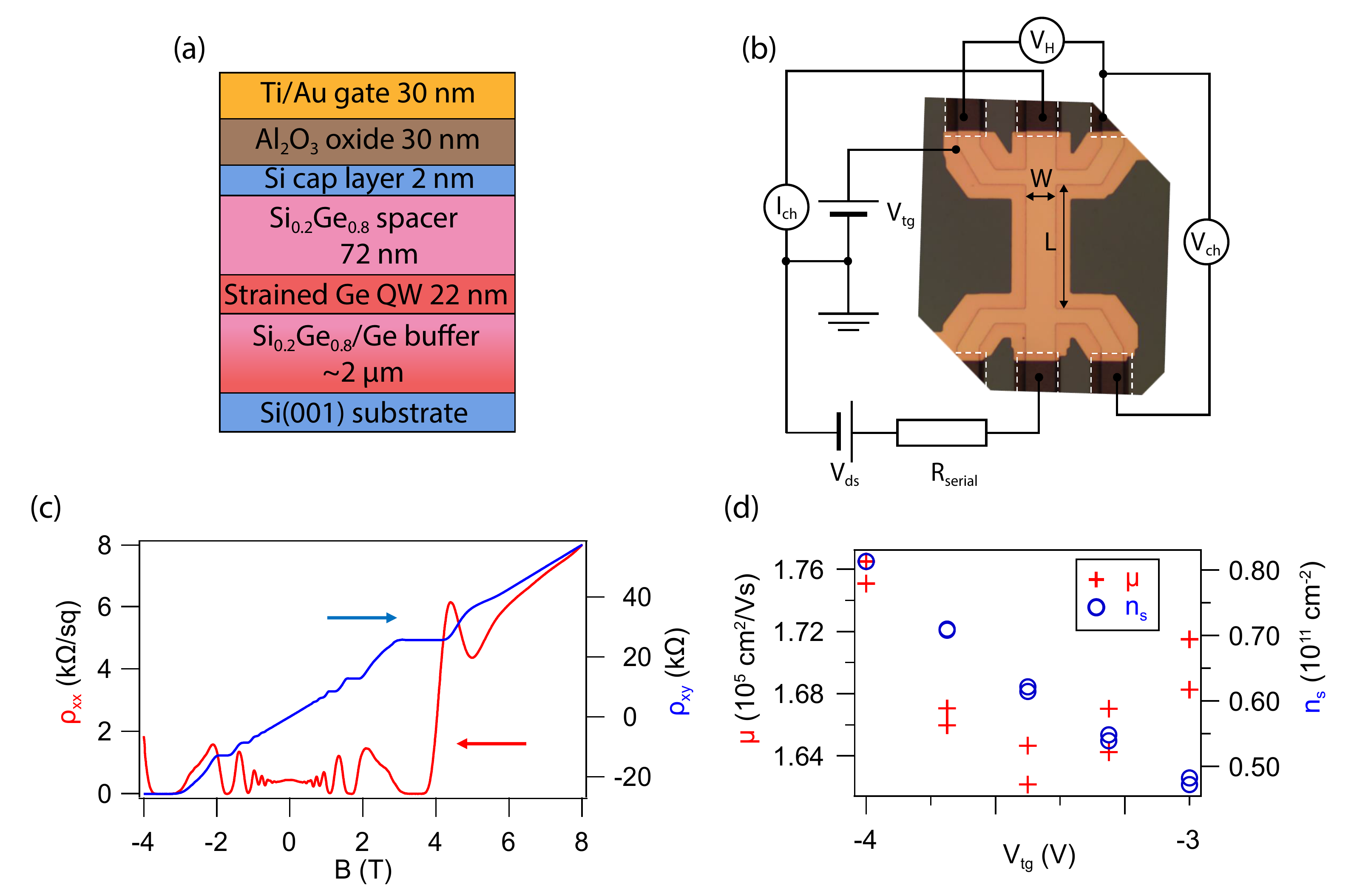}
	\caption{\label{Fig:Hallbar} (a) Schematic diagram of $\mathrm{ Ge/Si_{0.2}Ge_{0.8}}$ heterostructure with top gate. (b) Optical image of a gated Hall bar structure. White broken lines indicate six ohmic contacts. A top gate (yellow) overlaps each ohmic contacts and mesa structure. The mesa structure with a channel ($L = 80\ \si{\micro m}$ and $W = 20\ \si{\micro m}$) is seen through the top gate. The channel direction is [$\bar{1}10$]. A serial resistance $R_{serial} = 1\ \si{M \ohm}$ is connected to the channel. Constant bias voltage is applied and when the channel resistance is much lower than the $R_{serial}$ a constant current flows. The current through the channel $I_{ds}$, longitudinal voltage $V_{ch}$ and Hall voltage $V_{H}$ are measured at 300 mK as a function of gate voltage $V_{tg}$ or out-of-plane magnetic field $B$ and converted to longitudinal sheet resistivity $\rho_{xx} = V_{ch}/I_{ds}*W/L$ and Hall resistivity $\rho_{xy} = V_{H}/I_{ds}$. (c) Typical results of $\rho_{xx}$ and $\rho_{xy}$ vs $B$ for $V_{ds}$ = 100 mV and $V_{tg}$ = -4 V. Clear longitudinal resistivity oscillation (Shubnikov–de Haas effect) and Hall resistivity plateaus (quantum Hall effect) are observed (red and blue lines, respectively). At $B$ = 3 T, the filling factor $\nu = 1$. Around $B$ = 5 T, $\nu = 2/3$. (d) Hall density $n_s$ and hole mobility $\mu$ vs $V_{tg}$. $n_s$ is estimated from (classical) Hall effect in small magnetic fields and mobility $\mu$ is calculated for the relation $\mu = (en_s\rho_{xx})^{-1}$ at $B$ = 0, where $e$ is the electron charge. }
\end{figure}

The details of the heterostructure used in the present work are given in Fig. \ref{Fig:Hallbar} (a) \cite{Myronov2015Revealing,Morrison2016Complex}. To characterize the basic electronic properties of this  heterostructure, gated Hall-bar devices (Fig. \ref{Fig:Hallbar} (b)) were fabricated and measured at 0.3 K.  Representative measurements of longitudinal resistivity, $\rho_{xx}$, and Hall resistance, $\rho_{xy}$, are shown in Fig. \ref{Fig:Hallbar} (c). Shubnikov-de Haas (SdH) oscillations and quantum Hall plateaus are observed in $\rho_{xx}$ and $\rho_{xy}$, , respectively. The two-dimensional hole density, $n_s$, and the hole mobility, $\mu$, are plotted as a function of $V_{tg}$ in Fig. \ref{Fig:Hallbar} (d). In the shown $V_{tg}$ range, $n_s$ depends linearly on $V_{tg}$, reaching the largest value of $0.8 \times 10^{10}$ cm$^{-2}$ at the most negative $V_{tg}$. This is close to maximal hole density that could be achieved. In fact, by going to more negative $V_{tg}$, i.e. $V_{tg} < -4$ V, we encountered two types of problems: the accumulation of a parasite hole gas at the interface with the gate oxide \cite{lu2011upper, huang2014screening}, and  gate leakage. 

%%%%%%%%%%%%%%%%%%%%%%%% Estimation of Zeeman energy %%%%%%%%%%%%%%%%%%%%%%%%

\section{Measurement of the Zeeman energy}
\begin{figure}[p!]
	\centering
	\includegraphics[scale=0.4]{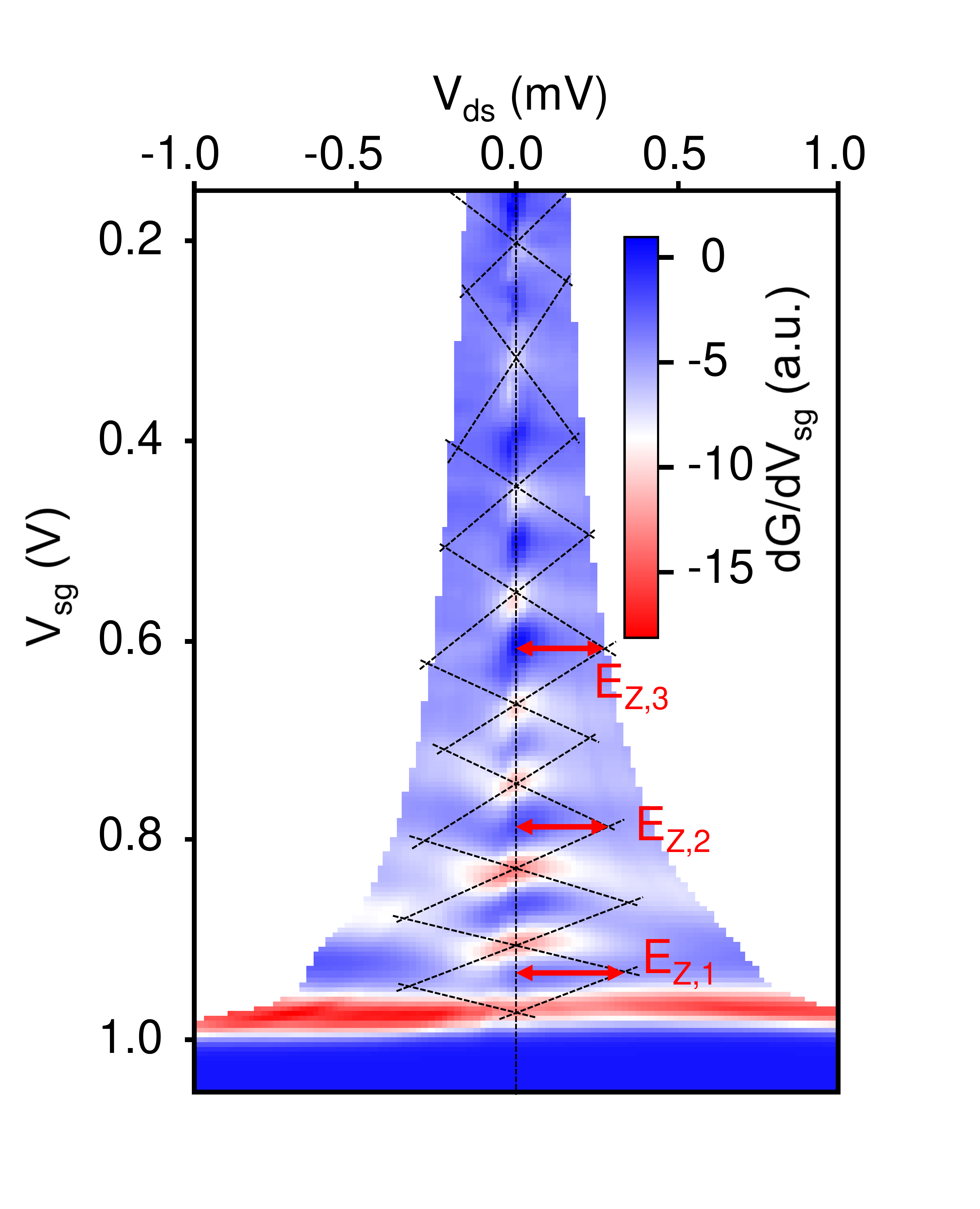}
	\caption{\label{Fig:Ezcalc} Color plot of $dG/dV_{sg}$ as a function of $V_{ds}$ and $V_{sg}$ at $B = 0.4$ T. The spanned  $V_{ds}$ range varies with $V_{sg}$. The dashed lines highlight  $dG/dV_{sg}$ ridges  forming a sequence of diamond-shape regions. The odd diamonds form from the spin splitting of the 1D subbands. Their half-width gives the Zeeman energy as indicated by the horizontal arrows. } 
\end{figure}

In this section we illustrate the procedure to measure Zeeman energy splittings. The color plot in Fig. \ref{Fig:Ezcalc} is a representative example of a $dG/dV_{sg}$ as a function of $V_{ds}$ and $V_{sg}$ at $B_z = 0.4$ T. 
The magnetic field is large enough to lift spin degeneracy. The diamond-shape blue regions centered around $V_{ds}=0\ \si{V}$ correspond to conductance plateaus at integer multiples of $e^2/h$. White/red lines bordering the diamonds define the edges of the plateaus. These lines are not always clearly visible. Dashed lines have been drawn to highlight their position. These lines correspond to aligning the energy of a subband edge with the Fermi energy of either the source or the drain lead.  
As a result, the apexes of the diamonds, defined by the crossings of consecutive  dashed lines, are located at a source-drain bias voltage equal to the energy spacing between consecutive subband edges. The horizontal half-widths of the odd diamonds provide a direct quantitative measurement of the Zeeman energies $E_{Z,n}$, as illustrated in 
Fig. \ref{Fig:Ezcalc}. 
The measurement accuracy can be conservatively estimated by varying the slope of the dashed lines until it becomes apparent that they no longer follow the  $dG/dV_{sg}$ ridges. Because the  $dG/dV_{sg}$ ridges happen to be  generally broad and sometimes even hard to identify, we end up with rather  large  measurement uncertainties. 

Besides providing access to the Zeeman splitting energies, the stability diagram of Fig.  \ref{Fig:Ezcalc} can be used to extract the gate lever-arm parameter, $\alpha$, which is the proportionality factor relating a gate voltage variation to the corresponding shift in the electrochemical potential in the 1D wire. In practice, for the n-th orbital subband $\alpha$ is obtained from the ratio between $E_{Z,n}$ and the height (measured along the $V_{sg}$ axis) of the 2n$-$1 diamond. We find that $\alpha$ decreases noticeably with n and, to a lower extent, it varies with  $\vec{B}$. 
For the case of  Fig. \ref{Fig:Ezcalc}  we find   $ \alpha \approx 5 \times 10^{-3} eV/V$ for n = 1,  $\alpha \approx 3.3 \times 10^{-3} eV/V$ for n=2, and   $\alpha \approx 2.3 \times 10^{-3} eV/V$ for n = 3. 

In the limit of vanishing $\vec{B}$, the odd diamonds shrink and disappear while the even diamonds grow. At $B=0$, the 2n diamond has a horizontal half-width set by the energy spacing $\Delta_{n,n+1}$ between the n-th and the (n+1)-th orbital subband.  We measure $\Delta_{1,2} \approx 0.65$ meV and $\Delta_{2,3} \approx 0.5$ meV.

\section{Data from device D3}
\begin{figure}[p!]
	\centering
	\includegraphics[width=1\linewidth]{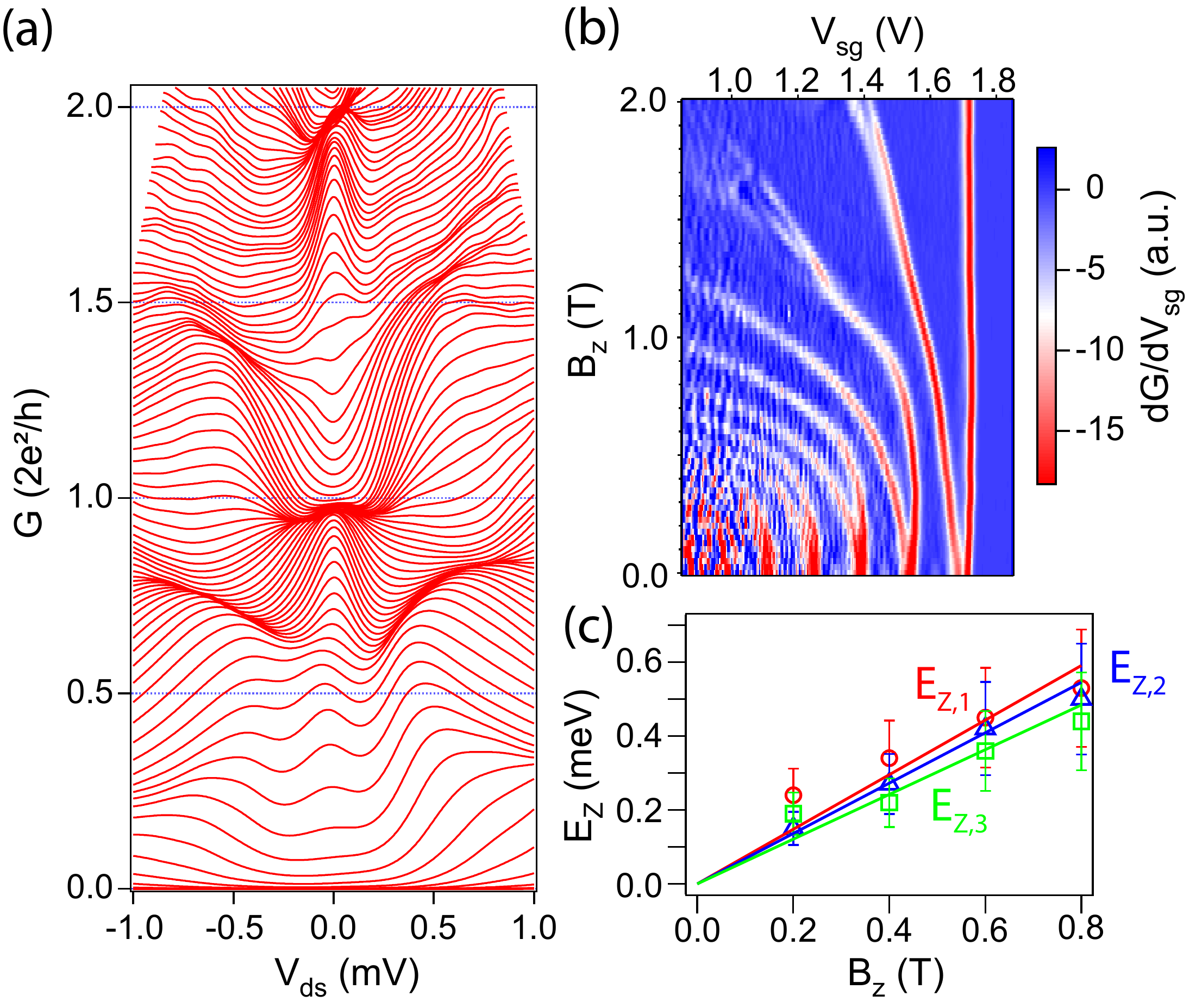}
	\caption{\label{Fig:Dev3} Experimental data for device D3.  (a) Differential conductance $G$ as a function of $V_{ds}$ at different $V_{sg}$ and  $B_z = 0$, (b) Linear transconductance $dG/dV_{sg}$ as a function of $V_{sg}$ and $B_{z}$, and (c) $E_{z}$ vs $B_{z}$. Large out-of-plane g factors are observed as in device D1 (see Table \ref{Table:gfactor} in the main text). }
\end{figure}

Figure \ref{Fig:Dev3} shows a set of data from a third device (D3) made from the same heterostructure. This device has the same gate layout as D1 as shown in Fig. \ref{Fig:QPC} (a) of the main text. It was measured with only one orientation of the applied magnetic field, perpendicular to the device plane ($z$-axis).  The procedure to correct for the series resistances, and the data analysis was the same as for the previous devices. The results are qualitatively and quantitatively similar to those from device D1. 
%$V_{tg}$ can be controlled in order to suppress oscillations on the plateaus due to Fabry-Pérot interference or backscattering, resulting flat plateaus. 
%The water-fall plot of Fig. \ref{Fig:Dev3} (a) shows traces of the differential conductance $G$ as a function of $V_{ds}$ at different $V_{sg}$. Fig. \ref{Fig:Dev3} (b) 
%\GRN{Its additional resistance $R_s$ at zero field depends less on bias than D1 ($R_s = 12.2 \pm 0.1 \ \si{k\Omega}$ for $|V_{ds}| < 2.5\ \si{mV}$), so constant $R_s$ correction was done for Fig. \ref{Fig:Dev3} to display clear zero bias anomaly. On the other hand, the bias dependence was taken into account for Zeeman energy calculation due to the fact that $R_s$  for high fields increases by 2$\ \si{k\Omega}$ at most. } The Zeeman energy of each subbnad is estimated also in D3 and plotted as a function of $B_z$ in Fig. \ref{Fig:Dev3} (c). \GRN{The $g$ factor of D3 is even higher than in D1 as shown in the main text (Table \ref{Table:gfactor}).}     

\providecommand{\latin}[1]{#1}
\makeatletter
\providecommand{\doi}
  {\begingroup\let\do\@makeother\dospecials
  \catcode`\{=1 \catcode`\}=2 \doi@aux}
\providecommand{\doi@aux}[1]{\endgroup\texttt{#1}}
\makeatother
\providecommand*\mcitethebibliography{\thebibliography}
\csname @ifundefined\endcsname{endmcitethebibliography}
  {\let\endmcitethebibliography\endthebibliography}{}

\end{document}